\pdfoutput=1
\documentclass[lettersize,journal]{IEEEtran}
\usepackage{amsmath,amsfonts}
\usepackage{algorithm}
\usepackage[caption=false,font=normalsize,labelfont=sf,textfont=sf]{subfig}
\usepackage{textcomp}
\usepackage{stfloats}
\usepackage{url}
\usepackage{verbatim}
\usepackage{graphicx}
\usepackage{algpseudocode}
\usepackage[cmintegrals]{newtxmath}
\usepackage{bm}
\usepackage{xcolor}
\def\BibTeX{{\rm B\kern-.05em{\sc i\kern-.025em b}\kern-.08em
    T\kern-.1667em\lower.7ex\hbox{E}\kern-.125emX}}
\usepackage{balance}

\begin{document}
\newcommand{\fullname}{Maximum Transmission Continuity}
\newcommand{\method}{MTC}
\newcommand{\UsetP}{P}
\newcommand{\UsetG}{G}
\newcommand{\UsetI}{I}
\newcommand{\chCBRS}{\mathbb{C}}
\newcommand{\chDP}{C}
\newcommand{\chPAL}{C_P}
\newcommand{\THS}{\gamma}
\newcommand{\pwrP}{Pwr_P}
\newcommand{\pwrG}{R} 
\newcommand{\numP}{|P|}
\newcommand{\numG}{|G|}
\newcommand{\utility}{U(t)}
\newcommand{\reqP}{\tilde{c}_{0,i}}
\newcommand{\reqG}{\tilde{c}_{1,j}}
\newcommand{\movP}{m_{0,i}}
\newcommand{\movG}{m_{1,j}}
\newcommand{\chP}{a_{0,i}}
\newcommand{\chG}{a_{1,j}}
\newcommand{\ch}{c}
\newcommand{\agII}{MSC}
\newcommand{\agIII}{TBSA}
\newcommand{\CA}{CA}
\newcommand{\fullagII}{Most Suitable Channel}
\newcommand{\fullagIII}{Type-Based Sequential Allocation}
\newcommand{\overTH}{\rho}
\newcommand{\stepfnc}{H}
\newcommand{\uu}{\omega}

\title{Reliable  Data Transmission through  Private CBRS Networks }

\author{
\IEEEauthorblockN{
Hsun-Yu Kuo, 
Szu-Yu Liu, 
Chin-Ya Huang, \IEEEmembership{Senior Member,~IEEE},
Yu-Chi Chen 
and 
Meng-Hua Xie
}
\thanks{H.-Y. Kuo is with the Bachelor Degree Program of Applied Science and Technology, National Taiwan University of Science and Technology, Taiwan.}
\thanks{S.-Y. Liu, C.-Y. Huang, Y.-C. Chen and M.-H. Xie  are with the Dept. of Electronic and Computer Engineering, National Taiwan University of Science and Technology, Taiwan.}
\thanks{Corresponding author: C.-Y. Huang (email: chinya@mail.ntust.edu.tw)}
}

\markboth{Journal of \LaTeX\ Class Files,~Vol.~14, No.~8, August~2021}%
{Shell \MakeLowercase{\textit{et al.}}: A Sample Article Using IEEEtran.cls for IEEE Journals}


\maketitle

\begin{abstract}
We consider the use of a domain proxy assisted private citizen  broadband radio service  (CBRS) network 
and propose a \fullname\ (\method) scheme to transmit Internet of Things (IoT) data reliably.  \method\ dynamically allocates available CBRS channels  to sustain the continuity of data transmission  without violating the channel access requirements.  \method\ allocates the granted CBRS channels according to the priority of each user, the instant channel access status, interference among users, and the fairness. The simulation results demonstrate the improvement in managing reliable IoT data transmission in the private CBRS network.
\end{abstract}


\section{Introduction} \label{sec:intro}

In the smart factory,  IoT sensors are often used to collect important data  which are further transmitted to a control center for the effective management and optimization of certain operations. In certain situations, sensor data may be time sensitive, or in some cases,  bandwidth intensive. For example,  collaborative machines require low latency data transmission to ensure the performance of their operations. More so, video  surveillance cameras need larger link bandwidths to adequately deliver sensor data \cite{John2022}. 

Interestingly, the Citizen broadband radio service (CBRS) band has the potential to  support  
 various  requirements of data transmission in the smart  factory \cite{W2022}. Following the specification \cite{forum}, the CBRS band can be released for further usage when the incumbent users are idle. In the CBRS network, a spectrum access system (SAS) manages the CBRS channel usages and grants the available channels to  the  citizen broadband radio service devices (CBSDs). The CBSDs are categorized as priority access license users (PALs) and general authorized access  users (GAAs). The PAL has higher priority than the GAA. Furthermore, for every CBSD interacting directly with the SAS, a domain proxy (DP) is introduced as an agent to request for CBRS channel access and to redistribute the granted channels to its served CBSDs.

In a smart factory,  several PALs and GAAs are deployed  to form a private CBRS network, where PALs deliver time sensitive applications, while GAAs take care of the other types of applications. These processes aim at simultaneously sustaining various types of data transmission in the network. However, the amount of traffic transmitted in the network might change dynamically due to limited  CBRS channels and the optional situations in the factory.  To address this drawback, the authors in \cite{Fair} propose mechanisms   that enable the SAS to allocate the CBRS channels to GAAs in order to sustain fairness among GAAs. In \cite{On(CBRS)}, geographic locations are applied to assist channel allocation among PAL and GAAs so that GAAs can sustain their network stability regardless of the large amount of channel access requests from the PALs. 
Further, algorithms are proposed to minimize the interference among GAAs and improve the network performance in \cite{Channel,SAS-Assisted}. Additionally, authors in \cite{W2022,Jai2021} propose mathematical approaches to optimize channel allocations for CBSDs aiming to satisfy the service requirements in the network.

In addition to CBRS networks, dynamic wireless channel allocation algorithms have been intensively studied in wireless networks. For example, in licensed shared access (LSA) system, the authors consider the regulation and propose a fair spectrum management algorithm to sustain the spectrum utilization and fairness in channel access \cite{Fair(LSA)}. 
In addition, the authors in \cite{Dynamic(SS)-3} use a channel quality indicator to assist dynamic channel allocation aiming to improve the network capacity.
Furthermore, channels are allocated to low priority users according to base on the arrival pattern in \cite{Channel(CR)_2},
and thus the interference of the high priority users can be reduced.

However, while most studies focus on allocating channels, the discontinuity in channel access during the  switch of the accessed channel is not considered. Specifically, frequent switching of the associated channels may result in the discontinuity of data transmission \cite{cont}.
Without any specific controls, the discontinuity in channel access would introduce additional  packet delay or loss, thereby limiting the operational performance of the smart factory.

In this paper, we develop a \fullname\ (\method) scheme that uses the DP  to maximize the continuity of CBRS channel access thereby  improving the channel utilization and the reliability of data transmission in the smart factory. With \method, the DP can dynamically allocate the available CBRS channels according to the priority of each CBSD,  the traffic waiting time in the CBSD and the transmission interference. 
As a result, the reliability of data transmission can be improved through maximizing the channel access continuity. 

\section{System Description and Problem Formulation} \label{sec:sys}
\begin{figure}[t] 
\centerline{\includegraphics[width=1.1\columnwidth]{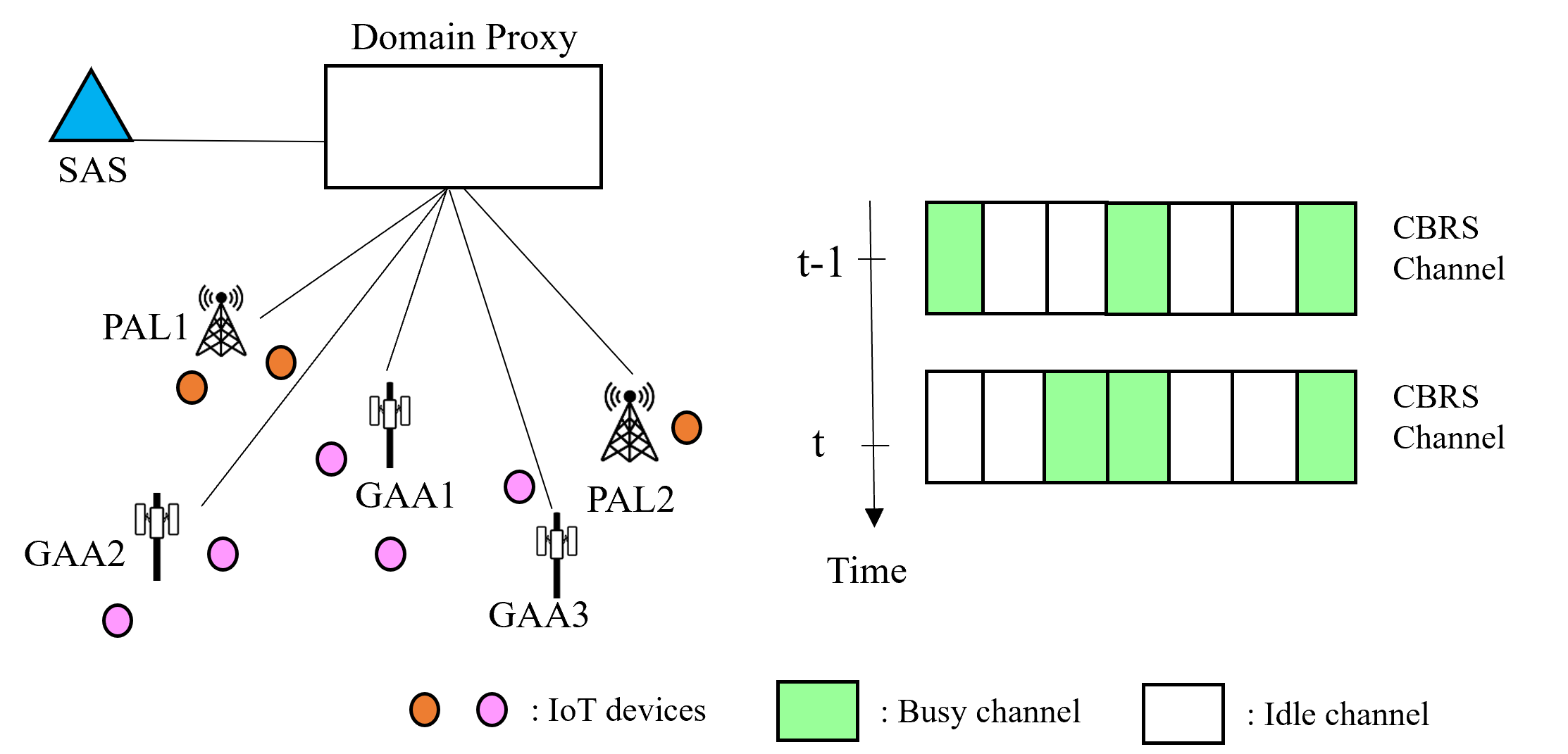}}
\caption{CBRS architecture.}
\label{fig:cbrs_arch}
\end{figure}
As illustrated in Fig. \ref{fig:cbrs_arch}. we assume  a private CBRS network is developed to support reliable data transmission in a smart factory. Following the CBRS protocol, the DP collects the channel requests from the CBSDs it serves every time slot and then  interacts with the SAS for CBRS channel access. Assuming  $\chCBRS$ channels, the SAS grants a set of channels, denoted as $\chDP$, to the DP based on the available channels and  the channel request from the DP.  Among the granted channels, PALs can only use a subset of channels, denoted as $\chPAL, \chPAL \subset \chDP$ while GAAs can conditionally access all granted channels according to the defined regulation. 

We denote the set of PALs and GAAs in the DP  as $\UsetP$ and $\UsetG$. Furthermore, we denote the requested  channels from PAL $i, i\in \UsetP$ and GAA $j, j \in \UsetG$ at the time slot $t$ as $\reqP(t)$ and $\reqG(t)$. Moreover,  
we denote 
$\chP^{\ch}(t)=1$ if \mbox{channel}\ s\ \mbox{is allocated to}\ \mbox{ PAL $i$} at time slot $t$. Otherwise, $\chP^{\ch}(t)=0$. 
Similarly, we denote $\chG^{\ch}(t)=1$ if \mbox{channel}\ \ch\ \mbox{is allocated to}\ \mbox{ GAA $j$} at time slot $t$. Otherwise, $\chG^{\ch}(t)=0$. 
Hence, 
    $\displaystyle \sum_{\ch\in \chPAL}\chP^{\ch}(t)\le \reqP 
    \mbox{}$ 
and      
    $\displaystyle  \sum_{\ch\in \chDP}\chG^{\ch}(t)\le \reqG 
    \mbox{.}$

According to the CBRS specification,  a GAA cannot access the channel $\ch, \ch\in \chDP$ if $\ch$ is allocated to a PAL, and thus we have   
    $\displaystyle  \sum_{i\in \UsetP}\chP^{\ch}(t) \sum_{j\in \UsetG}
    \chG^{\ch}(t)=0 
    \mbox{. }$
On the other hand, a channel $\ch, \ch\in \chDP$ can be simultaneously accessed by more than one GAAs as long as the interference  is below a threshold, $\THS$. Thus, 
the GAA $j$ can 
share the channel $\ch$ with other GAAs, when 
    $ \displaystyle  \sum_{j'\in \UsetG, j'\neq j}a_{1,j'}^{\ch}(t)\pwrG_{j, j'}\leq \THS 
    \mbox{, }$
where $\THS$ is the threshold and $\pwrG_{j,j'}$ 
is the interference caused GAA $j'$ to $j$.  
$\displaystyle \pwrG_{j,j'} = \frac{R_0}{(dx_{j}-dx_{j'})^2+(dy_{j} -dy_{j'})^2},$ 
where $R_0$ is assumed as the transmission power of each GAA and 
the geographic location of the GAA $j$ 
is denoted as $(dx_{j}, dy_{j})$.


Because of  the dynamic change in network conditions, CBSDs may dynamically request several channels. 
Furthermore, because PALs have a higher priority in accessing channels, a GAA might release its used CBRS channel to a PAL, resulting in transmission suspension or termination.
We denote  the number of accumulated  transmission suspensions or terminations caused by 
channel re-allocation for PAL $i$ and GAA $j$ at time slot $t$ as $\movP(t)$ and $\movG(t)$. 
The number of occurrences will increase by one if a CBSD must release a currently used channel even though the CBSD is still sending data. Specifically,\\
$\displaystyle \movP(t) = \movP(t-1) + \min(1,\sum_{\ch \in \chPAL}\max(\chP^{\ch}(t-1)-\chP^{\ch}(t), 0))$, 
$\displaystyle \movG(t) = \movG(t-1) + \min(1, \sum_{\ch \in \chDP }\max(\chG^{\ch}(t-1)-\chG^{\ch}(t), 0)).$

Traditionally, channels are allocated at every time slot $t$ aiming to satisfy all requests and service requirements without considering the channel usage at $t-1$. Under this condition, at $t$, a CBSD may use different channels from $t-1$ to transmit its data,  leading to transmission discontinuity and performance degradation while switching the channels. 
To maximize the service continuity by successively providing CBRS channel access opportunities to CBSDs without violating the CBRS regulation, we first define a utility, $\utility$, 
    \begin{equation}
    \displaystyle
    \utility = \sum_{i\in \UsetP}\frac{2}{\alpha^{\movP(t)}}+\sum_{j\in \UsetG}\frac{1}{\beta^{\movG(t)}}+2\sum_{\ch\in \chPAL}\sum_{i\in \UsetP}\chP^{\ch}(t) - \uu \cdot \overTH(t), 
    \label{eq:utility}
    \end{equation}
where $\alpha$, $\beta$ and $\uu$ are weightings to balance the channel allocation between PALs and GAAs. $\overTH(t)$ is the ratio indicating the number of CBSDs experiences larger interference  than $\THS$ among all CBSDs at time slot $t$, since the interference in the CBRS system should be  limited to $\THS$. Consequently, we formulate the achieved utility in response to the CBRS channel allocation in the DP as a maximization problem, as depicted in  Fig. \ref{fig:opt}.

\begin{figure}
\begin{align}
    & \hspace{-5mm} \text{max}\ \hspace{15mm}  U(t) 
    \nonumber 
    \\
    \text{s.t. }\ & \chP^{\ch}(t)=\{0,1\},\ 
    \forall i\in \UsetP,\ \ch\in \chPAL \label{eq:11a}\\
    & \chG^{\ch}(t)=\{0,1\},\ 
    \forall j\in \UsetG,\ \ch\in \chDP \label{eq:11b}\\
    & \hspace{-3mm} \sum_{j'\in G, j' \neq j} a_{1,j'}^{\ch}(t)\pwrG_{j, j'}\leq \THS,\ 
    \forall j\in \UsetG,\ \ch\in \chDP  \label{eq:11f}\\
    & \sum_{\ch\in \chPAL}\chP^{\ch}(t) \leq \reqP(t),\ 
    \forall i\in \UsetP \label{eq:11g}\\
    & \sum_{\ch\in \chDP}\chG^{\ch}(t) \leq \reqG(t),\ 
    \forall j\in \UsetG \label{eq:11h}\\
    &\sum_{i\in \UsetP}\chP^{\ch}(t) \sum_{j\in \UsetG}
    \chG^{\ch}(t)=0,\  
    \forall \ch\in \chDP  \label{eq:11i}\\
    & \sum_{i\in \UsetP}\chP^{\ch}(t)\leq 1,\ \ch\in \chPAL
    \label{eq:11j}\\
    & \movP(t) \leq \movP(t-1) + 1, \  
    \forall i \in \UsetP
    \label{eq:11l}\\
    & \movG(t) \leq \movG(t-1)+ 1, \ 
    \forall j \in \UsetG 
    \label{eq:11m}\\
    \nonumber
\end{align} \vspace{-10mm}
\caption{Problem formulation.}
\label{fig:opt}
\end{figure}

\section{\fullname\ (\method) Scheme} \label{sec:soln}

We propose a \fullname\ (\method) scheme to maximize the continuity of reliable data transmission 
where the DP dynamically allocates its CBRS channels, sent from the SAS to its served PALs and GAAs.
\method\ contains  \fullagIII\ (\agIII) algorithm  to enable the DP dynamically allocate channels based on the priority and the channel access status of each CBSD  to guarantee the service continuity by allocating each type of CBSD sequentially. Among the same type of CBSDs, we develop \fullagII\ (\agII) method to allocate the most accessible channel for each CBSD at time slot $t$, aiming to minimize the interference among GAAs and  the number of GAA  movements. Moreover, \fullagIII\ (\agIII) aims to maximize the channel access continuity by allocating channels according to the priority and the channel access status of each CBSD.

As illustrated in Algorithm \ref{alg:method}, \method\ first allocates channels to PALs, and then GAAs. Furthermore, \method\ allocates channels to PALs and GAAs using the same principle, \agIII,   by correctly setting the input of \agIII. The input of \agIII\ is summarized in Table \ref{tab:pca_input}.

\begin{table}[!h]
\centering
\caption{\agIII\ parameters setup for PAL and GAA.}
\begin{tabular}{|c|c|c|}
\hline
\agIII \ Parameter & Input of PAL & Input of GAA\\ \hline
$U$           & $\UsetP$ &  $\UsetG$ \\ \hline
$S$           & $\chPAL$ & $\chDP$ \\ \hline
$b_i^s$       & $\chP^s, s \in \chPAL$ & $\chG^s, s \in \chDP$ \\ \hline
$d_k$       & $\reqP, s \in \chPAL$ & $\reqG, s \in \chDP$ \\ \hline
$u$           & PAL & GAA \\ 
\hline
\end{tabular}
\label{tab:pca_input}
\end{table}

\begin{algorithm}
\caption{\fullname\ (\method)}
\label{alg:method}
\hspace*{\algorithmicindent}
\begin{algorithmic}[1]

\While{True}
    \State t $\gets$ t + 1
    \State $\chP(t)\gets$ \agIII(input for PAL)
    \Comment{TBSA with PAL input.}
    
    \State $\chG(t)\gets$ \agIII(input for GAA) 
    \Comment{TBSA with GAA input.}
\EndWhile
\end{algorithmic}
\end{algorithm}

\subsection{\agIII} 
Algorithm \ref{alg:agIII} defines the \agIII\ procedure allocating  channels at $t$ to maximize the objective function by considering the successive channel access of each CBSD. For CBSDs with the same priority, they are classified into three types according to their channel access status at $t-1$, `Allocated', `Moved', and `Remaining' CBSDs. 
An `Allocated' CBSD is the CBSD continuously using the same channel at $t-1$ and $t$. A `Moved' CBSD is the CBSD using a channel at $t-1$ that cannot access at $t$ because of the change in network conditions (Line 2-10). Hence, a `Moved' CBSD would be saved to an array $M$ for further channel allocation management using \agII\ (Line 11-17).  Finally, the `Remaining' CBSDs, the remainder of the `Allocated' and `Moved' CBSDs, are allocated last (Line 18-24). They are allocated following the principle, \agII, as the allocation of `Moved' CBSDs. 
With \agII, each CBSD keeps using the same  channel unchanged as much as possible to minimize the occurrence of discontinuity in channel access. Furthermore, \agII\ minimizes the interference among GAAs as much as possible when allocating multiple GAAs in the same channel.

\begin{algorithm}
\caption{\fullagIII\ (\agIII)}\label{alg:agIII}
\hspace*{\algorithmicindent} \textbf{Input:} {CBSD array $U$, and binary variable $b_k^s$, that can be 
$\chP^s$  
or $\chP^s$, and non-negative integer variable $d_k(t)$, that can be $\reqG^s(t)$ or $\reqP^s(t)$,  CBSD type $u \in \{PAL, GAA\}$, channel set $S$.}

\begin{algorithmic}[1]
\State $b_U^S(t) \gets 0$
\Comment{Initialize CBSD allocation status.}
\For{$\{(s, k)|b_k^s(t - 1) = 1, (s, k) \in S \times U$\}}
    \If{ $d_k(t) < \sum_{c \in \chDP} b_k^c(t)$ }
    \Comment{the channel requests of CBSD $k$ has not been satisfied.}
        \If{$\sum_{i \in \UsetP} \chP^s(t) = 0$}
        \Comment{CBSD $k$ is a `Allocated' CBSD.}
            \State $b_k^s(t) \gets 1$
        \Else
        \Comment{CBSD $k$ is a `Moved' CBSD.}
            \State Append $k$ to $M$
      \EndIf
    \EndIf
\EndFor
\For{$k \in M$}
\Comment{Allocate channels to all `Moved' CBSDs.}
    \If{$d_k(t) < \sum_{c \in \chDP} b_k^c(t)$}
        \State $e \gets (U, b_k^s, d_k(t), u, S)$
        \Comment{Save environment status to $e$}
        \State $s \gets$ MSC($e$, $k$)
        \State $b_k^s(t) \gets 1$
    \EndIf
\EndFor
\For{$k \in U$}
\Comment{Allocate channels to all `Remaining' CBSDs.}
    \For{$d_k(t)$ times}
        \State $e \gets (U, b_k^s, d_k(t), u, S)$
        \Comment{Save environment status to $e$}
        \State $s \gets$  MSC($e$, $k$)
        \State $b_k^s(t) \gets 1$
    \EndFor
\EndFor
\State Return $\{b_k^s(t)| k \in U, s \in \chDP\}$
\end{algorithmic}
\end{algorithm}


\subsection{\agII}
Algorithm \ref{alg:agII} presents the \agII\ choosing the optimal channel for each CBSD at $t$  to enhance GAA transmission continuity while minimizing the interference among GAAs in the same channel. When allocating a channel to a PAL, the channel with the least number of GAAs is chosen, because PAL has higher priority accessing the channel $s, s\in\chPAL$ (Line 3-4). In this way, the least number of GAAs are moved during this allocation. Furthermore, GAAs can share the same channel under the Condition (4). Therefore, to minimize the interference  fitting the Condition (4) among GAAs, the channel $s$ is selected by (\ref{eq:cim}) from $A'$ (Line 5-10), where $A$ is the channel space, 

\begin{equation}
\begin{aligned}
A' = \{s'|\sum_{i \in \UsetP} \chP^{s'}(t) = 0, \sum_{j'\in G, j' \neq j} \Big(a_{1,j'}^{s'}(t)\pwrG_{j,j'} \Big) \leq \THS, s' \in \chDP \}
\end{aligned} \label{eq:cspace}
\end{equation}

\begin{equation}
\begin{aligned}
s= \operatorname*{argmin}_{s' \in A'} \sum_{j'\in G, j' \neq j} a_{1,j'}^{s'}(t)\pwrG_{j, j'}.
\end{aligned} \label{eq:cim}
\end{equation}

\begin{algorithm}
\caption{\fullagII\ (\agII)}\label{alg:agII}
\hspace*{\algorithmicindent} \textbf{Input:} {CBSD array $U$, and binary variable $b_k^s(t)$, that can be $\chP^s(t)$ or $\chG^s(t)$, CBSD type $u \in \{PAL, GAA\}$, channel set $S$ and the allocation channel $k$.}

\begin{algorithmic}[1]
\State $A \gets \{s'| \sum_{i \in \UsetP} \chP^{s'}(t) = 0, s' \in S\}$
\Comment{Put all available channels into the set A.}
\State $s \gets \emptyset$
\If{$u = PAL$}
    \Comment{The CBSD is a PAL}
    \State $s = \operatorname*{argmin}\limits_{c\in A}(\sum_{j\in \UsetG} \chG^c(t-1))$
\Else
    \Comment{The CBSD is a GAA}
    \State $A' \gets \{s'|\sum_{j'\in G, j' \neq k} a_{1,j'}^{s'}(t)\pwrG_{k, j'} \leq \THS, s' \in A\}$
    \If{$A' \neq \emptyset$}
        \State $s= \operatorname*{argmin}\limits_{c\in A'}(\sum_{j'\in G, j' \neq k} a_{1,j'}^{c}(t)\pwrG_{k, j'})$
    \EndIf
\EndIf
\State Return $s$
\end{algorithmic}
\end{algorithm}

\subsection{Example}

\begin{figure}[t] 
\centerline{\includegraphics[width=1.1\columnwidth]{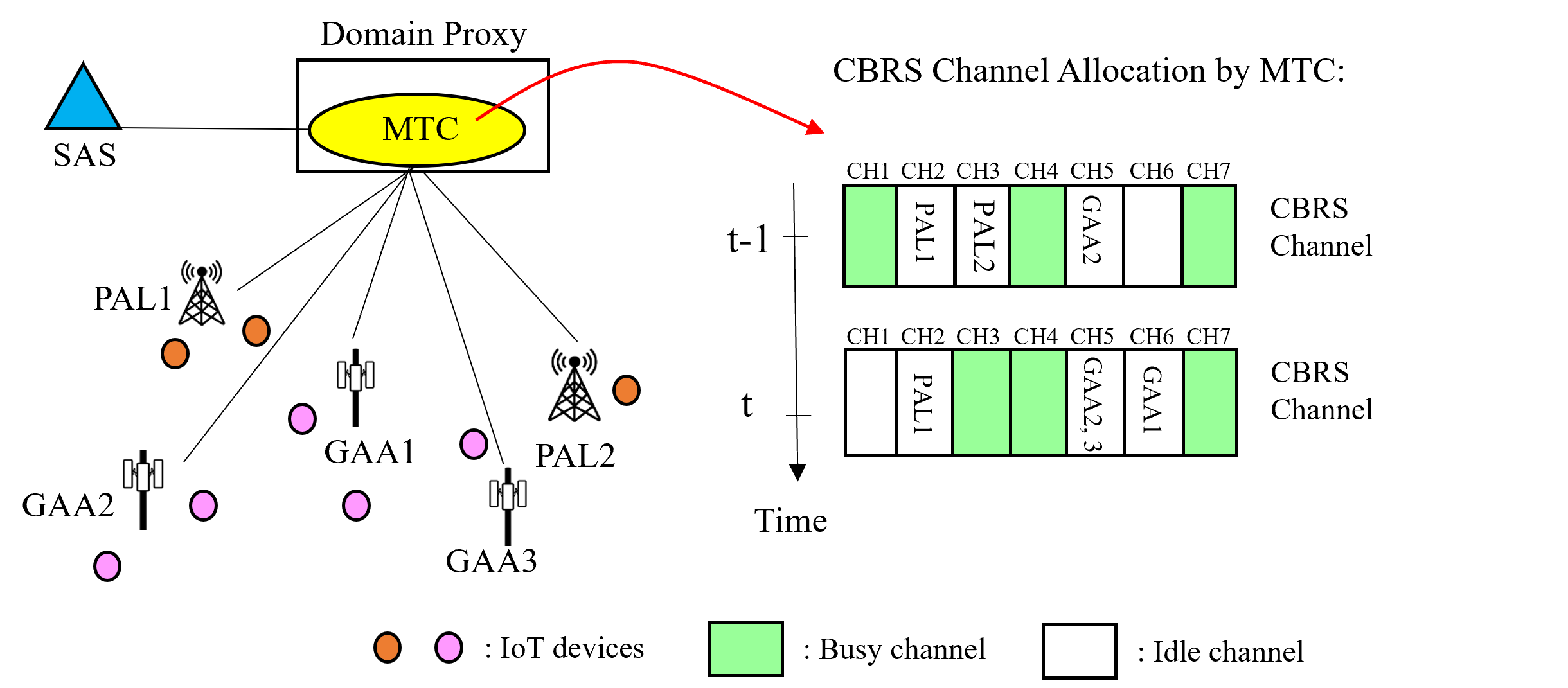}}
\caption{Illustration of CBRS channel allocation by MTC.}
\label{fig:cbrs_allocated}
\end{figure}

Fig. \ref{fig:cbrs_allocated} presents an example of channel allocation by \method. When CBRS channels are available, we assume CH1- CH3 are allocated for PALs while CH4-CH7 are shared by PALs and GAAs. At t-1, CH2, CH3, CH5 and CH6 are available and PAL1, PAL2 and GAA2 individually request a channel. According to \agIII, \method\ allocates each a channel. At t, CH1, CH2, CH5 and CH6 are available for CBSDs while PAL 2 leaves the network and GAA 1 as well as GAA 3 join the network and each requests a channel. Since at t-1, PAL1 and GAA2 have allocated CH2 and CH5, respectively, according to \agIII, they consecutively use their assigned channels. On the other hand, based on \agIII\ and \method, GAA3 shares CH5 with GAA2 and CH6 is allocated to GAA1 because GAA1 is geographically located between GAA2 and GAA3 which would introduce higher interference if it shares the channel with GAA2 or GAA3.

\section{Evaluation Results} \label{sec:eval}

We consider that CBSDs are normally distributed within a 4 km$\times$ 4 km geographic area. 
Among the 50 time slots of an experiment, the PAL channel requests  increase by 1 and 2 with 15 \% and 5 \% probabilities within the first 25 time slots and  1 with 12.5 \% within the second 25 time slots. Furthermore, those decrease by 1 with 12.5 \% probability within the first 25 time slots and  1 and 2 with 15 \% and 5 \% probability within the second 25 time slots. Furthermore, the increment of GAA channel requests is uniformly distributed over $(0,5)$ and $(0,2)$ within the first and second 25 time slots. The decrement of GAA channel requests is uniformly distributed over $(0,2)$ and $(0,5)$ within the first and second 25 time slots.   
Furthermore, we set $\chDP$ and $\chPAL$ as uniform distributions between 7 and 10 and 4 and 7. We also set the transmission power of a CBSD to 10 watts, $\THS=0.01$, $\alpha=2$, $\beta = 4$
 and $\uu = 10$.

Three strategies are introduced for performance comparison. First, we modify the solution in \cite{Channel} and create a channel allocation (\CA) scheme to compare our proposed \method. 
Second, we also employ first in first serve (FIFS) approach, which allocates the channels according to the arrival sequence of CBSDs.
Third, we adopt the priority based (PB) approach that first allocates the channels to the higher priority CBSDs. Both FIFS and PB allocate only one  channel to one request.

\begin{figure}[t] 
\centerline{\includegraphics[width=1.1\columnwidth]{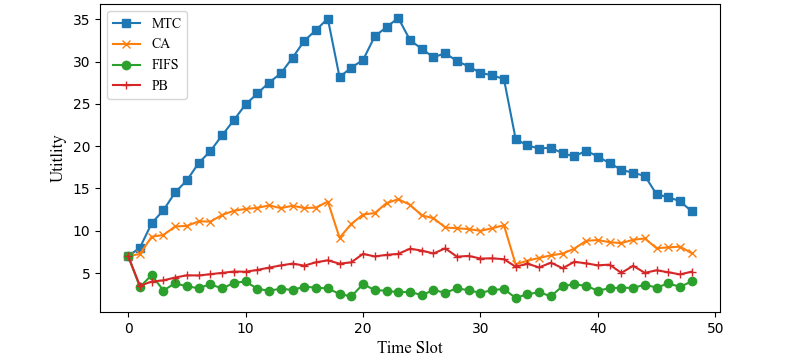}}
\caption{Utility over time.}
\label{fig:utility}
\end{figure}

Fig. \ref{fig:utility} illustrates the corresponding utility under the different strategies. \method\ outperforms others because  it considers not only the priority  but also  the continuity in channel access during channel allocation. \CA\ aims at satisfying all channel requests without considering the movement of CBSDs accessing the channels in the previous time slot and interference among GAAs leading to limited performance. PB and FIFS allocate channels based on the priority of CBSDs and the order of requests, obtaining low utilities.

\begin{figure}[t] 
\centerline{\includegraphics[width=1.1\columnwidth]{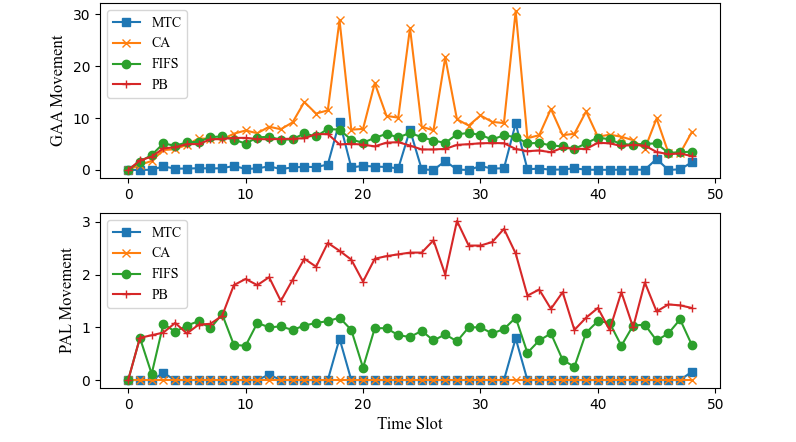}}
\caption{Number of CBSD movements over time.}
\label{fig:move}
\end{figure}

Fig. \ref{fig:move} presents the number of CBSD movements over time. The number of movements of \method\ is the smallest because in \method, the CBSDs can continuously access the same  channels with higher probability  which improves the utility. However, other schemes do not consider the current channel usage status for the subsequent allocation management so that their movements are higher, leading to low utility.

Table \ref{tab:overall} summarizes the average ``Unallocated PAL Requests'', ``Unallocated GAA Requests'' and the ``Interference Violation Ratio''. 
FIFS and PB do not allocate multiple GAAs in the same channel, and thus, as  confirmed by``Unallocated PAL Requests'' and ``Unallocated GAA Requests'' present, some CBSD requests are not satisfied instantly. Under these circumstances, their achieved utilities are low. 
Furthermore, \CA\ focuses on satisfying all channel requests, but does not consider the interference requirement, (\ref{eq:11f}). Consequently, some transmissions violate the interference requirement, resulting in low transmission performance and utility. 
\method\  allocates channels depending  on the priority and channel access status of each CBSD but  also allocates as many GAAs as possible without violating the interference constraint, $\THS$. Consequently, the ``Unallocated PAL and GAA Requests'' and the ``Interference Violation Ratio'' are small contributing to improved network performance.

\begin{table}
    \caption{Performance under different strategies.}
    \centering
    \begin{tabular}{|c|c|c|c|}
        \hline
        &   Unallocated  & Unallocated   & Interference  \\
        &  PAL Requests &  GAA Requests  &  Violation Ratio \\
        \hline
        \method\ 
        & 0.21 & 2.13 & 0\\ 
        \hline
        CA 
        & 0.21 & 0  & 0.25\\
        \hline
        FIFS 
        & 1.39 & 14.72  & 0\\
        \hline
        PB  
        & 0.21 & 16 & 0\\
        \hline 
    \end{tabular}
    \label{tab:overall}
\end{table}

\section{Conclusion} \label{sec:summary}

In this paper, we formulate a utility function and then propose the \fullname\ (\method) scheme to dynamically allocate the available CBRS channels in order to promote the data transmission continuity of IoTs. 
The evaluation results show the proposed \method\
can effectively reduce the number of CBSD movements as well as ensure the priority in channel access. In the future, we will integrate the proposed \method\ into a real DP assisted CBRS private network 
for performance evaluation and enhancement.

\section*{Acknowledgment}
This work was partially supported by the Taiwan Building Technology Center from The Featured Areas Research
Center Program within the framework of the Higher Education Sprout Project by the Ministry of Education, Taiwan, R.O.C. and National Science and Technology Council, Taiwan, R.O.C. under Grant no. NSTC 108-2221-E-011-058-MY3  and 111-2221-E-011-092.

\normalsize
\bibliography{ref.bib}
\bibliographystyle{ieeetr}

\end{document}